\documentclass[letterpaper,twocolumn,showpacs,pra,aps]{revtex4}
\usepackage[colorlinks,urlcolor=blue]{hyperref}
\usepackage{amsmath}  
\usepackage{amsfonts} 
\usepackage{graphicx} 
\usepackage{amssymb}

\begin{document} 
\title{The physical meaning of the magnetic scalar potential,\\ and how to use it to design an electromagnet}
\author{C. B. Crawford}
\affiliation{Department of Physics and Astronomy, University of Kentucky, Lexington, KY 40506-0055, USA}
\date{Nov 26, 2020}
\begin{abstract}
The magnetic scalar potential can be used to design electromagnets accurately and efficiently.  I will describe a practical construction algorithm: the prescribed field in a ``Target'' region (constrained only by Maxwell's equations) specifies boundary conditions which uniquely determine the potential in the surrounding field ``Return'' region.  As required by the physical representation of the magnetic scalar potential, the conducting elements of the resulting coil are directed along equipotentials on the surface of each region, at equal increments of the potential.  I give some examples and comments on the limits of precision of the constructed field.
\end{abstract}
\pacs{
07.55.Db, 
41.20.Gz, 
03.50.De 
}
\maketitle

\section{Making a magnetic field with a coil}
The practical use of magnetism often requires the design of a current-carrying collection of wires (a ``coil'') that creates a specified magnetic field.  Some examples are:
a) NMR requires a very uniform field, so that a spin precesses at the same rate no matter where it may be in the experimental volume~\cite{Cates,Lemdiasov}; for example searches for the electric dipole moment of the neutron~\cite{Khriplovich} require gradients smaller than 1~pT/cm~\cite{Pignol,Ahmed}.
b) Transporting polarized atoms or neutrons~\cite{Maldonado} requires a holding field that varies slowly with position so that the spin can follow the field adiabatically.
c) Active magnetic shielding can be constructed to cancel the field of a fixed source~\cite{Wyszynski:2017qvj}.
d) Guiding particle beams while maintaining stability of the trajectory and pulse duration requires a field with specified magnetic moments~\cite{Green,Walstrom90}.
The characteristic problem requires a magnetic field $\vec B (\vec r)$ of specified form within a ``Target'' volume, to a stated tolerance.  Of course, the desired field will have to obey the conditions
\begin{equation}
\label{Ampere}
\vec \nabla \times \vec H = 0      
\end{equation}
(since it would be undesirable to have currents inside the region of study and the scalar potential must be well defined) and 
\begin{equation}
\label{Gauss}
\vec \nabla \cdot \vec B = 0,     
\end{equation}
where the fields $\vec B$ and $\vec H$ are related by a constitutive equation, for example a proportionality $\vec B = \mu \vec H$.  In much of what follows, it will be assumed that $\mu = \mu_0$ everywhere, although this is not a requirement.

In general there must be a surrounding ``Return'' region, since components of the magnetic field normal to the surface of the Target cannot be canceled.   It is usually desirable (but not required for this method) that the Return is a bounded region and that no field extends outside these regions, both to prevent the created field from interfering with other devices, and to avoid having magnetizable objects outside the coil becoming part of the magnetic system (see Figure 1).  The current-carrying wires are on surfaces at the boundary between the Target and the Return, on the exterior surface of the Return, and possibly on other surfaces dividing the Target and the Return into disjoint regions.   
\begin{figure}
\includegraphics[width=1.0\columnwidth,keepaspectratio]{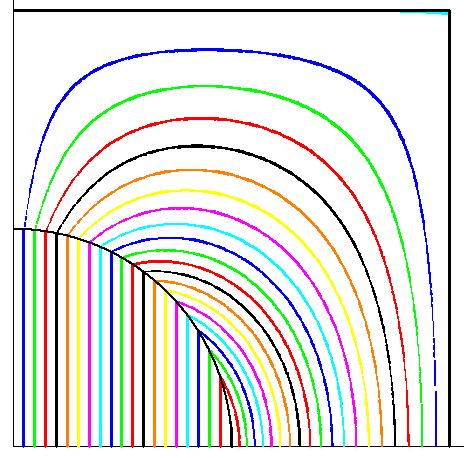}
\caption {(Color online).  Cross section of one quadrant of a magnetic coil that creates a uniform transverse field in a tube of circular cross-section (the Target), enclosed in a Return with square outer boundary.  The field is tangent to the lines, and the separation between lines indicates the field strength.  There are currents flowing perpendicular to the page on the circle and square borders, corresponding to the discontinuity in the parallel component of the field.}
\end{figure}

The traditional workflow for designing coils involves starting with a basic design, calculating the magnetic field, and iterating changes to the windings to converge on the desired field configuration.   Many techniques exist in the literature~\cite{Turner93} for streamlining this process, for example: a) through the use of basis current loops to optimize the target field numerically~\cite{Turner86,Lemdiasov2,Xu,Thompson,Gluck,Rawlik}, either solving the linear system of equations for the currents, or optimizing nonlinear coil parameters.  The progression from a single coil to the Helmholtz~\cite{Helmholtz,Purcell} and Maxwell~\cite{Maxwell} arrangements is the ancestor of this approach, and has been extended to higher harmonics~\cite{Garrett}.  Since the field and currents are related by linear equations, this method is sure to work, but the relationship between the desired field and the required coil is only as clear as one's understanding of an integral equation involving a tensor Green function.  Surface and bulk current distributions have been described as fictitious magnetization densities~\cite{Dasgupta, Walstrom91, Walstrom94, Liu} which can take an arbitrary form in the limit of infinitesimal dipoles.  Various optimiation procedures include variational principles~\cite{Thompson,Liu} and linear programming~\cite{Xu}.

In problems with either real or fictitious magnetization density, the net pole density plays the role of electric charge in an analogous boundary value problem to that of the electric field~\cite{Chechurin, Ciric, Dasgupta, Walstrom02}.  For real current distributions stream functions specify the resulting discontinuity of the scalar potential~\cite{Haus, Mayergoyz, Lemdiasov}, which has also been applied to eddy currents~\cite{Krahenbuhl} and the analysis of fields~\cite{Muniz}.  \cite{Gross} showed that any volume containing surface currents can be divided into topologically simple regions in which the scalar potential is well-defined.  This technique has been applied in cylindrical, spherical, and planar geometries to calculate surface current windings of coils with minimized higher order multipoles~\cite{Walstrom90, Walstrom02}.

In this paper I will describe a construction algorithm based on the physical interpretation of the magnetic scalar potential that can create any field $\vec B (\vec r)$ satisfying (\ref{Ampere}) and (\ref{Gauss}) within any set of Target regions, and with only weak physical constraints on the boundaries and necessary Return regions deriving from (\ref{Ampere}) and (\ref{Gauss}), though practical and technical considerations such as ease of construction, tolerance requirements, and power management add further constraints.   These will be discussed subsequently.  The particular focus is on extremely well characterized fields, with relative variation as little as $10^{-6}$ from the design field, with tight geometric constraints on where currents may be placed.  This method inverts the process of magnet design, starting from the desired field and calculating the required windings on specific surfaces to produce this field.

Because this technique involves the design of surface current coils, it is applicable to low-field precision coils, with low enough current densities that they may be formed as a single layer of conducting strips or wires, still dissipating an acceptable amount of power.    This limits the field to about 10~G (though much larger for superconductors), but the generalization to multiple layers is straightforward.  
This approach follows an intuitive physical interpretation of the magnetic scalar potential, which allows visualization of any electromagnetic coil and the currents required to create it.

\section{ The magnetic scalar potential, and the constructive algorithm based on it}

\subsection{ Theory}
Inside a region with no currents, Eq. (\ref{Ampere}) implies that we can construct a magnetic scalar potential $U$ such that
\begin{equation}
\label{ ScalarPot}
\vec H(\vec r) = - \vec \nabla U (\vec r)
\end{equation}
Within the Target, the field $\vec H$ is completely specified (this is the starting point for the problem to be solved), and thus we are given $U$ for that region.    The fields in the Return must have the same normal component of B at the interface with the Target, and vanishing normal component of $\vec B$ at the exterior surfaces for a hermetic coil, or other constraints, such as zero current, on parts of the surface.  In order to explain how these are connected and determine the coil windings needed to relate them, it is convenient to introduce (conceptually) a thin layer between adjacent regions, and ascribe windings to the surface of each region separately.  The property of this ``Transition'' region will be that the $H$ field is normal to the two surfaces; there is no tangential field.  Thus the magnetic scalar potential on the Transition side of the surface does not vary as one moves along the surface and thus is a constant at the surfaces of the Target and the Return, which can be set to zero by making the Transition arbitrary thin; thus, we can avoid having to make a detailed study of it.

Imagine a set of equipotential surfaces that correspond to values of the scalar potential that differ by multiples of a small constant amount $\Delta U$.  The neighboring surfaces cannot intersect, and they separate the boundary of the Target into ribbons that form closed loops  (Figure 2).   The magnetic field usually has a component parallel to the boundary, while within the transition region it does not, by construction; then there is a current sheet on the ribbon.   Its magnitude is determined according to the integral form of Ampere's law\begin{equation}
\label{ Ampere}
\oint \vec H(\vec r) \cdot  d \vec r = I         
\end{equation}
For a path near the surface that encloses a ribbon, the part of the path inside the Target gives the difference in the magnetic scalar potential; since the field is normal to the surface in the transition region, there is no contribution.   Thus the current in the ribbon is given by  $I = -\Delta U$~\cite{Fye,Haus,Walstrom90,Warnick} (in S.I. units), where the $\Delta$ implies the difference between the two edges of the ribbon.   Choosing  $\hat n$ to be the outward directed normal to the surface, this describes a surface current density 
\begin{equation}
\label{CurrentDensity}
 -\hat n \times \vec H = \hat n \times \vec \nabla U  = \vec K(\vec r)     
\end{equation}
on the boundary that flows along lines of constant magnetic potential. This is an expression of the continuity boundary condition arising from (\ref{Ampere}) by replacing $\nabla$ with $\hat n$, $\vec H$ with $\Delta \vec H$ (the difference in $\vec H$ between the outer (Transition) and inner (Target) sides of the surface), and $\vec J$ with $\vec K$.  The second continuity equation deriving from (\ref{Gauss}) is
\begin{equation}
\Delta \hat n \cdot \vec B = 0
\end{equation}
or equivalently
\begin{equation}
\label{normal}  
\Delta \partial_n(\mu U) = 0
\end{equation} 
Equations (\ref{CurrentDensity}) and (\ref{normal}) are discrete versions of  $\nabla \times \vec H = \vec J$ and $\nabla \cdot \vec B = 0$.
\begin{figure}
\includegraphics[width=1.0\columnwidth,keepaspectratio]{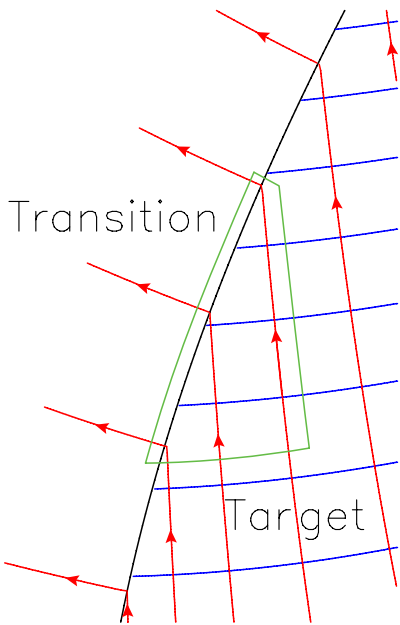}
\caption {(Color online).  The field and scalar potential near a boundary.    The boundary (black line) separates the Target from the Transition region.   The equipotentials of the magnetic scalar potential (blue lines) divide the boundary into segments, which are parts of the ribbons around the region.   The field lines (red)  are perpendicular to the equipotentials.   The scalar potential in the Transition region is constant along the boundary and can be set to zero; it is related to that within the Target only by having the same normal component of $\vec B$ at the boundary.       The green polygon is a path of integration for Ampere's law, such as that described in the text.}
\end{figure}
The fields in the Return can also be described by a magnetic scalar potential $U'$ which satisfies the Laplace equation.   At the boundaries the normal component of $\vec B$ is specified (Neumann boundary conditions): at the boundary with the Target, the normal component of $\vec B$ is given by the normal component of $\vec B$ in the Target) from (\ref{normal});  at the exterior boundary the normal component of $\vec B$ is zero, since there is no field outside the Return.  Dirichlet boundary conditions may be substituted on a portion of either boundary to constrain a region with specified surface currents (for example, no current); however this will result in slight modifications of the specified fields in the Target or leakage fields out of the Return.  This boundary value problem determines the magnetic scalar potential up to an immaterial additive constant; it can be found numerically using Finite Element Analysis \cite{FEA}.   As in the case of the Target, the equipotentials of $U'$  slice up the boundary of the Return into closed ribbons, which carry current $I' = -\Delta  U'$ to cancel the tangential field just outside the boundary.
The introduction of the Transition region separates the boundary conditions for the Target from those for the Return, allowing for the simple physical significance of the scalar potential, which is local to each region.   The way $U$ and $U'$ are defined directly determines the surface current density on the boundaries, so that these prescribe the coil windings in the actual device.   It should be noted that the potentials in the various regions are not continuous across the boundary,  and are related only by (\ref{CurrentDensity}) and (\ref{normal}).  The utility of winding separate currents around each region is that the distribution of branching at the junction path of 3 or more regions is automatically handled.  In some cases, it is advantageous to combine the Target and Return portions of current along adjacent sides of the boundary into a single set of windings calculated by discretizing the combined potential $U-U'$; however in these cases, junction currents must be handed by manually splitting the currents~\cite{Hayes} or double-winding on a separate boundary, such as the end-caps of a single-wound long double-cylinder coil~\cite{Maldonado}.
\begin{figure}
\includegraphics[width=1.0\columnwidth,keepaspectratio]{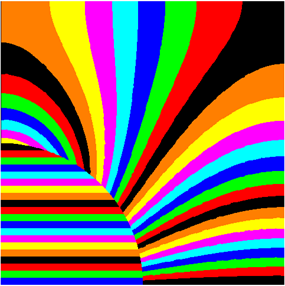}
\caption {(Color online) The magnetic scalar potential corresponding to Figure 1.  The colored bands represent regions across which the scalar potential differs by a constant amount $\Delta.$   There are current sheets on the circular and square boundaries directed along the axis of the cylinder (perpendicular to the view), with equal current within each colored zone at the surface.}
\end{figure}

The essential point is that the magnetic scalar potential $U$ has a direct physical meaning:  its series of equipotential contours on the boundary of a region represents the path and current of source windings needed to generate the field $\vec H = - \vec \nabla U$ within that region.  In particular, the current flows along the isocontours of $U$, and the total current flowing between any two isocontours of values $U_1$ and $U_2$ is equal to $I = U_1 - U_2$ as in stream functions.  These windings and their corresponding contours of the scalar potential ``fence in'' the field by terminating its transverse component $\hat n \times \vec H$ at the boundary, leaving only the normal component to pass through the Transition to the next region.   Fencing the fields in all regions sharing magnetic flux lines establishes an electromagnetic coil that not only guides the magnetic field, but also generates it.  However, if the windings from one region are omitted then the field will not be normal in its adacent Transitions, invalidating the interpretation of the scalar potential in the surrounding regions.

The result of this construction algorithm is a coil assembly that can create a magnetic field of arbitrary form (consistent with Maxwell's equations (\ref{Ampere})  and (\ref{Gauss}) within the Target and with no field escaping beyond the Return, in the form of a surface current density flowing in closed loops along the equipotential lines.  The solution is unique, once the shape of the Return is chosen.  Although it is possible to replace the two current sheets on either side of the Transition by one, leaving them separated avoids the problem of getting the surface currents to split up correctly at the junction of three or more boundaries \cite{Maldonado}.

The magnetic scalar potential describes the free current analog of bound surface currents, which effectively circulate around the boundary of regions of magnetic material when the magnetization field $\vec M(\vec r)$ is irrotational.  This accounts for the similarities in bound and unbound design techniques.  For example, a Dirichlet boundary condition in the magnetic scalar potential and its corresponding surface currents can be represented by a magnetic dipole layer~\cite{Walstrom02}, just as an electric dipole layer generates a discontinuity in the electric potential.

A shortcut for handling the Return, eliminating the need for calculation, is to replace the Return with a shell of a ferromagnetic material ($\mu \rightarrow \infty$) thick enough that $B << B_{sat}$, which satisfies the same condition $\Delta \vec H_t=0$ as imposed in the Transition~\cite{Fye}.  Inside the ferromagnetic Return, windings are only required along equipotentials of the Target region.  If the entire boundary is perpendicular to Target field then the inner windings also vanish almost everywhere except between the north and south pole tips where magnetic flux enters and exits the Target, respectively.

The scalar potential method for determining the winding geometry of a coil extends naturally to regions with spatially varying magnetization, either expressed as local permeability $\mu(\vec r)$ and/or permanent magnetization $\vec M(\vec r)$, by replacing Laplace's equation with the scalar elliptic equation
\begin{equation}
\label{Elliptic}
- \vec \nabla \cdot \mu \vec \nabla U = -\mu_0 \nabla \cdot \vec M_0 = \rho_{M},
\end{equation}
derived from the full constitutive equation, where $M_0$ is the portion of magnetization not included in $\mu$.  This equation is the analog of electrostatic problems, with a magnetic charge source term $\rho_{M}$ and a flux source $\sigma_{M}=-\hat n\cdot\vec M_{0}$ on the boundary.  The magnetic scalar potential $U$ still represents the free currents which must be added to generate the Target field.

Interior surface currents $\vec K$ or discontinuities in $\mu$ necessitate partitioning the volume into separate regions $\mathcal{R}_i$, with potentials $U_i$ related by the continuity equations (\ref{CurrentDensity}) and (\ref{normal}).  In some cases it may be desirable or necessary to separate the Target or the Return into several compartments with a current layer separating each.  This might allow a larger field for the same current, by avoiding surfaces where the magnetic field has a large change in tangential component.  In the case of the Target, where the internal fields are completely specified, the boundary conditions require continuity of the normal component of $\vec B$, and determing the current layer $\vec K$ from the discontinuity in the parallel component of $\vec H$.  In the case of the Return, any current sheet on the interface between different regions of the Return is undetermined and must be specified in the continuity boundary conditions (5) and (6) between the two scalar potentials, which are solved for as one boundary value problem.

\subsection { Practice}
For a practical coil, the current will be discretized into either a) a 3-dimensional printed circuit board~\cite{Avery,Koss} with solid ribbons of conductor between pairs of equipotential contours, such that the current density in each ribbon approximates a step function, as determined by Ohm's law; or b) wires wound along each equipotential contour carrying all of the current assigned between it and the contour of the next wire, which has the advantage of a a precisely known current density (a delta function at each wire) which does not depend on geometry and the local thickness of conductive traces in the first method. 
The discretization from a continuous current density $K(\vec r)$ to either a step function of constant $K$ across each ribbon or a delta function of constant $I$ in each wire causes an alteration of the field inside the Target near the boundary that falls off exponentially as one goes away from the boundary, with a scale length given by $D/2 \pi$ (where $D$ is the wire spacing)\footnote{For example, uniformly spaced wires on the $xy$-plane give a surface current density
$$ K(x) = I \sum_{n} \delta(\frac{nx}{D}) = \frac {I}{D} [1 + 2 \sum_ {n} \cos (\frac{2 \pi nx}{D})].$$
 The corresponding scalar potential is 
$$ U(x,y,z) = \frac {I}{D} [x+2 \sum_{n} \sin (\frac{2\pi nx}{D}) \exp(- \frac{2 \pi nz}{D})], $$
 where the first term represents the field of a uniform current sheet and the rest is the effect of discretization. }.
In regions of nonlinear variation of $K$ across each ribbon, the field can die off much slower than exponential.  In practice, one minimizes the residual fields by tuning the actual values of each equipotential used for each winding.  For example, previous design placed the wire at the barycenter of current in each ribbon~\cite{Bidinosti}.

The coil as constructed by this algorithm consists of independent closed loops, each carrying the same current.   These can be connected together in series, allowing the entire boundary to be wound with a single wire, by winding around the highest $U$-contour first, then crossing over to the next contour with a short leader, winding that contour, and so on to the end.   After winding the last contour, the wire is traced back along each leader to cancel the current that is not part of the real solution \cite{Walstrom}.   The two ends of the wire are now near the same point, to be connected to the current supply power by a twisted pair or coaxial line.   This process of joining the loops into a series must be done for each region of nonzero field to obtain a complete coil producing the desired field in the Target.

This strategy of following the equipotential line and joining to the next contour at a step avoids a pair of problems.  Evening out the displacement (as is done in a helically wound solenoid) introduces a new surface current parallel to the magnetic field.    This will give an error field in the interior, except for rotationally symmetric windings.   In any case, the current has to be in a complete circuit;  having wound around the Target, there has to be a connection back to the power supply.   The field due to one wire circling back will depend on the wire spacing $D$ and the distance $R$ of the return wire from the target  $\delta H  \approx  I/R$, so that $R\delta H/H \approx  D/R \approx 10 ^ {-3}$.   Having the return current right over the steps joining the turns allows one problem to solve the other.

The cancellation of the field of the crossing leaders is only first order, replacing the $H_1 \approx I/2 \pi r$  field of the string of leaders by a line of  dipoles with a field $H_2 \approx  I \delta /r^{2}$ where $\delta$ is the average separation between the leaders and the current return line ($\delta$ can be smaller than the average spacing $D$ between wires, but is comparable to it).   To put these estimates in context, the field in the Target is $H_0 \approx  I/D$; for a coarsely wound coil the wire spacing is of order $10^{-2}$ the smallest dimension of the coil, so that $H_1 \approx 10^{-2} H_0$ and $H_2 \approx 10^{-4} H_0$.
In what follows I will continue to describe the coil as if it were the collection of loops, while asserting that the practical implementation as a series circuit is a negligible change.

\section{Some examples}

The technique described above is applicable to arbitrary geometries and target fields.  In the following section, we describe highly symmetric geometry, for which the winding geometry is readily apparent by visualizing the magnetic equipotentials.  A few nontrivial extensions are described through their corresponding boundary value problems.

\subsection{Solenoids}
A uniform magnetic field has regularly spaced, parallel, planar equipotentials.    An infinite solenoid of arbitrary fixed cross-section will have uniform axial field $H=I/D$ if the coil is wound with current $I$ around the perimeter of cross-sectional slices with spacing $D$ normal to the field lines.
Segments of straight solenoid can be joined to produce a ``bent solenoid'' meeting at an angle $2 \alpha$.  The two cylinders meet at a curve in a plane whose normal is tilted by the angle $\alpha$ from the field lines in either segment.  Close to the bend, the equipotentials run along the plane of intersection perpendicular to the field lines on either side.  The ovelapping windings on the interface of both segments generates a planar current density  $K =  \frac{2I}{D} \sin \alpha$, which kinks the field, reversing the tangential component of $\vec H$ between neighboring sections~\cite{Koss}.  This gives a straightforward geometric interpretation of a detailed analysis result based on Biot-Savart integration~\cite{Abele}.

A finite solenoid with uniform field inside can be constructed by adding current-carrying end caps.  The current distribution on the end caps is calculated by solving the scalar potential in the infinite region outside the solenoid with constant flux boundary conditions at each end face.   
The field outside the cylindrical surface of the solenoid is in the opposite direction as the constant field inside the solenoid, and therefore the outside windings are in the same direction as the inside ones.   The equipotentials outside are closest together near the ends, indicating a higher current density at the ends to prevent fringing inside the finite-length solenoid.

\subsection{Cos-theta coils}
A uniform magnetic field perpendicular to the axis of a circular cylinder (with radius $A$ and axis $\hat z$) has a magnetic scalar potential $U = - H_0 x = - H_0 \rho \cos \phi$, giving rise to the name, with $\theta$ instead of $\phi$.   This can be created with inner windings that are equally spaced in $x$ with current $I = H_0 \Delta x$.  The field outside the cylinder is described by the potential $U' = - H_0 (A^2/\rho) \cos \phi  = -H_0 x (A/\rho)^2$.    Thus the outer windings have exactly the same current as the inside windings, for double the total surface current.
That design has an infinite Return; for a Return in the form of a larger concentric cylinder with radius $B$, the magnetic scalar potential is $U' = - H_0 A^2 (\rho/B^2 + 1/\rho) \cos \phi$ has zero flux escaping the outer cylinder.  The form of the modified field is shown in Figure 4.  The infinite cylinder solution is well known~\cite{Bidinosti}, and can also be calculated by considering a second cos-theta coil at radius $B$ with the opposite magnetic moment $H_A A^2 = -H_B B^2 = (B^{-2}+A^{-2})^{-1} H_0$, where $H_{A,B}$ are the interior fields of $\cos\theta$ coils of radii $A$, $B$, with current densities $K_{A,B}=2H_{A,B}\cos\phi$, respectively, such that the total field is $H_0$ at the origin.  However, the physical interpretation of the magnetic scalar potential immediately yields the winding configuration required to truncate the double-$\cos\theta$ coil to finite length while preserving the $z$-symmetry: both the cylindrical sections and the end caps are wound on equally spaced contours of $U$ and $U'$.  Ascribing separate boundaries and windings to each region simplifies the winding patterns at edges where three or more regions meet compared to single windings, in which current must be diverted between multiple surfaces where they meet.  The design of a transverse adiabatic radio frequency spin rotator based the double-cos-theta coil designed in this manner is described in \cite{Hayes}.

\begin{figure}
\includegraphics[width=1.0\columnwidth,keepaspectratio]{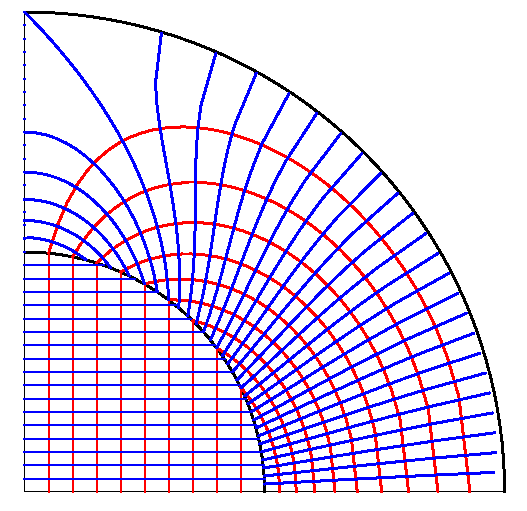}
  \caption{(Color online). Field lines (red) and equipotentials (blue) for the
    $\cos (\theta)$ coil.  The potential is discontinuous across the
    boundaries, corresponding to the current sheet flowing on the surfaces.}
\end{figure}

As with solenoids, transverse coils can be wound on any interior and exterior cross section, for example a square inside of an octagon, and the field strength can taper along $z$~\cite{Maldonado}.  The case of a square outer boundary is shown in Figures 1 and 3.

\subsection{Active Magnetic Shielding}
This is the spatial inverse of the magnet coil.   We are given a surrounding field due to external flux sources and nearby magnetic materials.   One can solve the boundary value problem for $U$ in the neighborhood outside the shielded room that either: a) is consistent with the sources and has vanishing normal component of $\vec B$ at the surface; or b) has no external sources except the opposite flux through the surface as that caused by the external sources.  The external region may contain permeable materials as described above.  The resulting contours of constant potential on the surface of the proposed shielding coil tell how to wind it most effectively and with best use of power.   The coil can be wound fairly coarsely, since the field penetration will exponentially decrease with a length scale given by the wire spacing.   Separate coils can be calculated independently for a few discrete sources of flux, or two or three basis functions of a uniform background field of variable direction to tune out individual interferences.

\subsection{ Use of magnetizable materials and superconductors}
When the magnetic field in the Target is already normal to the surface, the magnetic potential is a constant on the surface and all surface current occurs between the regions on the boundary with inward and outward magnetic flux.   This technique used to design pole tips along equipotentials of the desired field~\cite{Marble} for electromagnets.   By using a highly permeable ($\mu \rightarrow \infty$) ``flux-return yoke'' outside the Target with windings, the generated magnetic flux distributes along the pole tips, in the desired configuration.  Conversely, when the desired magnetic field is parallel to the surface,  the surface currents confine the flux inside the target.   Here a perfect diamagnet ($\mu = 0$---a superconductor) can be used in  place of windings to satisfy the boundary conditions; supercurrents will serve to fence the flux inside.

\section{ Limitations} 
The attainable accuracy of the resulting field in a realization is determined by the deviations from the assumptions of the theory.   Here is a brief discussion of the leading concerns:

\paragraph{Geometrical accuracy of the construction of the coil}  Misplacement of a wire from its ideal position by $\delta x$ over a span $Y$ creates an error field of dipolar form with magnitude $\delta H/H\approx \delta x DY/R^3$ at distance $R$ from the wire ($D$ is the spacing between wires.  The extra factor $D/R$ is the fractional contribution of one wire to the total field).   Since $D$ and  $\delta   x$ will typically be of order $10^{-3} R$,  this might be ignorable, but a systematic displacement of a region will give coherently adding errors.  For example,  shifting the windings of an infinite solenoid of radius $A$ by a distance $\delta x$ in a finite region of length $L$  gives an effect equivalent to an extra turn carrying current $I \delta x/D$ at the two ends of the region, and gives rise to an error field of magnitude $I \delta x/DA$.   The wire form must be accurately constructed, and care be taken that the wire follows the intended path, especially at corners.

\paragraph{Effects of the conversion of independent loops to a series-connected coil}  As already noted, this gives a dipolar field of order  $\delta   H/H\approx  10 ^{-4}$.    Weaving the step and the return current lines to give alternating sign dipoles might decrease this by another factor of $10^{-2}$ but would not be practical in many construction techniques.   Another option is to double wind the coil, spiraling up and then down again; this cancels the effects of the conversion to a helical winding in a uniform way, again at the cost of complexity.

The current has to be in a complete circuit;  having wound around the Target, there has to be a connection back to the power supply.   The field due to one wire circling back will depend on the wire spacing $D$ and the distance $R$ of the return wire from the target  $\delta   H \approx   I/R$, so that $\delta   H/H \approx  D/R\approx  10 ^{-3}.$   The schemes discussed in the previous paragraph avoid this error.      

\paragraph{Discretization of the current sheet}  As discussed above, this is exponentially suppressed, so that the disturbance falls off as $\exp(-2\pi z/D)$ as one moves away from the surface.  For wire spacing $D$, the error field at distance $2 D$ is of order $\exp(-4 \pi) = 3\times10^{-6}$; the exponential suppression assumes that discrete wires differ from the surface current density required by the algorithm only by a periodic correction.   The discreteness becomes important where the wires are widely spaced, because the scale length becomes large, or when the width varies, so that the exponential form is invalid.  For example, the winding on a sphere to give a uniform field inside has the wires widely spaced towards the poles, with the result that the field is inaccurately produced there\cite{Nouri}  

\paragraph{Maximum field} For normal metals the largest attainable field is set by the limit on surface heating, which in term depends on the resistivity of the coil material and the surface current density.  There are related by $K \approx H$; the Joule heating/area is $P \approx \rho K^{2}/D \approx H^{2}/D$, where $\rho$ is the resistivity and $D$ is the thickness of the winding layer (effective thickness of the wires used).   Thus for 1 $W/cm^{2}$, $D=10^{-3}$ m, and $\rho \approx 10^{-8} \Omega m$, $H < 300$~Oe.
For a superconducting winding, the limit is set by the critical field.

\section{Acknowledgments}
The author would like to thank Prof. Joseph P. Straley for fruitful conversations.  This work was supported in part by the U.S. Department of Energy, Office of Nuclear Physics under contracts DE-SC0008107 and DE-SC0014622, and by the National Science Foundation under award number PHY-0855584.

\end{document}